\documentclass[12pt]{iopart}
\usepackage[utf8]{inputenc}
\usepackage[T1]{fontenc}
\usepackage{xspace}
\expandafter\let\csname equation*\endcsname=\relax
\expandafter\let\csname endequation*\endcsname=\relax
\usepackage{amsmath,amssymb,amsfonts,amsthm}

\catcode`,\active

\catcode`\,12

\newcommand*\pFqskip{8mu}
\catcode`,\active
\newcommand*\pFq{\begingroup
        \catcode`\,\active
        \def ,{\mskip\pFqskip\relax}%
        \dopFq
}
\catcode`\,12
\def\dopFq#1#2#3#4#5{%
        {}_{#1}F_{#2}\biggl[\genfrac..{0pt}{}{#3}{#4};#5\biggr]%
        \endgroup
}
\newcommand{\Keywords}[1]{\par\noindent
{\footnotesize{\textbf{Keywords}\/}: #1}}
\newcommand{\nombrespacs}[1]{\par\noindent
{\small{PACS numbers\/}: #1}}
\newcommand{\class}[1]{\par\noindent
{\small{AMS classification scheme numbers\/}: #1}}
\renewcommand{\phi}{\varphi}
\newcommand{\wt}{\widetilde}
\newcommand{\ket}[1]{|#1\rangle\xspace}
\newcommand{\kket}[2]{|#1\rangle_{#2}\xspace}
\newcommand{\bra}[1]{\langle #1|\xspace}

\newcommand{\bbraket}[4]{{}_{#1}\langle #2|#3\rangle_{#4}\xspace}
\newcommand{\BBraket}[5]{{}_{#1}\bra{#2}#3\ket{#4}_{#5}\xspace}
\numberwithin{equation}{section}
\begin{document}
\title[Multivariate Meixner polynomials \& $SO(d,1)$ representations]{The multivariate Meixner polynomials as matrix elements of $SO(d,1)$ representations on oscillator states}
\author{Vincent X. Genest}
\ead{genestvi@crm.umontreal.ca}
\address{Centre de recherches math\'ematiques, Universit\'e de Montr\'eal, Montr\'eal, Qu\'ebec, Canada, H3C 3J7}
\author{Hiroshi Miki}
\ead{hmiki@mail.doshisha.ac.jp}
\address{Department of Electronics, Faculty of Science and Engineering, Doshisha University, Kyotanabe City, Kyoto 610 0394, Japan}
\author{Luc Vinet}
\ead{luc.vinet@umontreal.ca}
\address{Centre de recherches math\'ematiques, Universit\'e de Montr\'eal, Montr\'eal, Qu\'ebec, Canada, H3C 3J7}
\author{Alexei Zhedanov}
\ead{zhedanov@yahoo.com}
\address{Donetsk Institute for Physics and Technology, Donetsk 83114, Ukraine}
\begin{abstract}
The multivariate Meixner polynomials are shown to arise as matrix elements of unitary representations of the $SO(d,1)$ group on oscillator states. These polynomials depend on $d$ discrete variables and are orthogonal with respect to the negative multinomial distribution. The emphasis is put on the bivariate case for which the  $SO(2,1)$ connection is used to derive the main properties of the polynomials: orthogonality relation, raising/lowering relations, generating function, recurrence relations and difference equations as well as explicit expressions in terms of standard (univariate) Krawtchouk and Meixner polynomials. It is explained how these results generalize directly to $d$ variables. 
\\
\vfill

\nombrespacs{03.65.Fd, 02.20.-a}\\
\class{06B15, 33C50, 22E46}
\end{abstract}

\pagebreak
\section{Introduction}
The objective of this paper is to provide a group theoretical interpretation of the multivariate Meixner polynomials and to show how their properties naturally follow from this picture. The Meixner polynomials in $d$ variables will be shown to arise as matrix elements of the reducible unitary representations of the pseudo rotation group $SO(d,1)$ on oscillator states. For simplicity, the emphasis will be placed on the $d=2$ case, where the bivariate Meixner polynomials occur as matrix elements of $SO(2,1)$ representations. The extension of these results to an arbitrary finite number of variables is direct and shall be presented at the end of the paper.

The standard Meixner polynomials of single discrete variable were defined by Meixner \cite{Meixner-1934} in 1934  as polynomials orthogonal on the negative binomial distribution
\begin{align*}
w^{(\beta)}(x)=\frac{(\beta)_{x}}{x!}(1-c)^{\beta}c^{x},\qquad x=0,1,\ldots
\end{align*}
with $\beta>0$, $0<c<1$ and where $(\beta)_{x}=(\beta)(\beta+1)\cdots (\beta+x-1)$ stands for the Pochhammer symbol \cite{Koekoek-2010}. These polynomials possess a number of interesting features such as a self duality property, an explicit expression in terms of the Gauss hypergeometric function, a second order difference equation, etc. \cite{Koekoek-2010} and have found numerous applications in combinatorics \cite{Bergeron-1990-09,Foata-1983-12}, stochastic processes \cite{Karlin-1957-11,Karlin-1958}, probability theory \cite{Borodin-2006-12,Johansson-2001} and mathematical physics \cite{Floreanini-1993-09,Johansson-2000-02}. They also enjoy an algebraic interpretation as they arise in the matrix elements of unitary irreducible representations of the $SU(1,1)$ group \cite{Vilenkin-1991}.

The multivariate Meixner polynomials were first identified by Griffiths in 1975. In his paper \cite{Griffiths-1975-06}, Griffiths defined the polynomials through a generating function and gave a proof of their orthogonality with respect to a multivariate generalization of the negative binomial distribution. The same Meixner polynomials were considered by Iliev in \cite{Iliev-2012}. Using generating function arguments, Iliev established the bispectrality of these polynomials, i.e. he gave the recurrence relations and difference equations they satisfy, and also gave an explicit expression for them in terms of Gel'fand-Aomoto hypergeometric series. In both cases, the multivariate Meixner polynomials came in as generalizations of the multivariate Krawtchouk polynomials \cite{Griffiths-1971-04,Grunbaum-2011-12}, which are multivariate polynomials orthogonal on the multinomial distribution (see \cite{Diaconis-2013-09,Genest-2013-06} and references therein for additional background on the Krawtchouk polynomials).

Recently in \cite{Genest-2013-06,Genest-2013-07-2}, a group theoretical interpretation of the multivariate Krawtchouk polynomials was found in the framework of the $d+1$-dimensional isotropic quantum harmonic oscillator model. More specifically, it was shown that the multivariate Krawtchouk polynomials in $d$ variables arise as matrix elements of reducible unitary representations of the rotation group $SO(d+1)$ on the eigenstates of the $(d+1)$-dimensional isotropic harmonic oscillator. The group theoretical setting allowed to recover in a simple fashion all known properties of the polynomials and led to addition formulas as well as to an explicit expression in terms of standard (univariate) Krawtchouk polynomials. The approach moreover permitted to determine that the multivariate generalization of the Krawtchouk polynomials introduced by Tratnik in \cite{Tratnik-1991-04} are special cases of the general ones associated to $SO(d+1)$.

The algebraic interpretation of the multivariate Meixner polynomials proposed here in terms of the pseudo-orthogonal group $SO(d,1)$ is in a similar spirit. The relevant unitary reducible representations of $SO(d,1)$ will be defined on the eigensubspaces of a $SU(d,1)$-invariant bilinear expression in the creation/annihilation operators of $d+1$ independent harmonic oscillators. This embedding of $\mathfrak{so}(d,1)$ in the Weyl algebra will allow for simple derivations of the known properties of the polynomials and will also lead to new formulas stemming from the group theoretical context. This will provide a cogent underpinning of the multivariate Meixner polynomials.

The paper is organized as follows. In Section 2, the reducible unitary representations of $SO(2,1)$ on the eigenspaces of a $SU(2,1)$-invariant bilinear expression in the creation/annihilation operators of three independent harmonic oscillators are constructed. In Section 3, it is shown that the matrix elements of these representations are given in terms of polynomials in two discrete variables that are orthogonal on the negative trinomial distribution. The unitarity of the representation is used in section 4 to obtain the duality property satisfied by the polynomials. In Section 5, a generating function is obtained and is identified with that of the multivariate Meixner polynomials. In Section 6, the recurrence relations and the difference equations satisfied by the multivariate Meixner polynomials are derived. In Section 7, the matrix elements of natural one-parameter subgroups of $SO(2,1)$ are related to the standard  Meixner and Krawtchouk polynomials. In section 8, addition formulas and a number of special cases of interest related to possible parametrizations of $SO(2,1)$ elements are discussed. In particular, these considerations lead to explicit expressions of the multivariate Meixner polynomials in terms of standard (univariate) Meixner and Krawtchouk polynomials. In Section 9, the analysis presented in details for the bivariate case is extended to an arbitrary number of variables, thus establishing that the $d$-variable Meixner polynomials occur as matrix elements of reducible unitary representations of the $SO(d,1)$ group. A short conclusion follows.

\section{Representations of $SO(2,1)$ on oscillator states}
In this section, the reducible $SO(2,1)$ representations on oscillator states that shall be used throughout the paper are defined. These representations will be specified on the infinite-dimensional eigensubspaces of a bilinear expression in the creation/annihilation operators of three independent harmonic oscillators.

Let $a_i$, $a_i^{\dagger}$, $i=1,2,3$ be the generators of the Weyl algebra satisfying the commutation relations
\begin{align*}
[a_i,a_k]=0,\qquad [a_i^{\dagger},a_{k}^{\dagger}]=0,\qquad [a_{i},a_{k}^{\dagger}]=\delta_{ik}.
\end{align*}
This algebra has a standard representation on the vectors
\begin{align}
\label{Basis}
\ket{n_1,n_2,n_3}\equiv \ket{n_1}\otimes \ket{n_2}\otimes \ket{n_3},\quad n_1,n_2,n_3=0,1,\ldots,
\end{align}
and is defined by the following actions on the factors of the direct product:
\begin{align}
\label{Action-2}
a_i\ket{n_i}=\sqrt{n_i}\ket{n_i-1},\quad a_{i}^{\dagger}\ket{n_i}=\sqrt{n_i+1}\ket{n_i+1}.
\end{align}
Consider the Hermitian operator
\begin{align}
\label{Hamiltonian}
H=a_{1}^{\dagger}a_{1}+a_{2}^{\dagger}a_{2}-a_{3}^{\dagger}a_{3}.
\end{align}
It is seen that \eqref{Hamiltonian} differs by a sign from the standard Hamiltonian of the three-dimensional isotropic harmonic oscillator. As opposed to the latter, $H$ does not have a positive definite spectrum. Indeed, it is easily seen that $H$ is diagonal on the  oscillator states \eqref{Basis} with eigenvalues $E=n_1+n_2-n_3$, that is
\begin{align*}
H\ket{n_1,n_2,n_3}=E\ket{n_1,n_2,n_3}.
\end{align*}
It is obvious from the expression \eqref{Hamiltonian} that $H$ is invariant under $SU(2,1)$ transformations. We introduce the set of orthonormal basis vectors
\begin{align}
\label{DEF-BASIS}
\kket{m,n}{\beta}=\ket{m,n,m+n+\beta-1},\qquad m,n=0,1,\ldots
\end{align}
where $\beta\geqslant 1$ takes integer values. The vectors \eqref{DEF-BASIS} span the infinite-dimensional eigenspace associated to the eigenvalue $E=1-\beta$ of $H$. These vectors support an irreducible representation of the $SU(2,1)$ group generated by the symmetries of $H$ which are of the form $a_{i}^{\dagger}a_{j}$, $a_{3}^{\dagger}a_{3}$, $a_{i}a_{3}$ and $a_{i}^{\dagger}a_{3}^{\dagger}$ for $i,j=1,2$. In the following, we shall concentrate on the subgroup $SO(2,1)\subset SU(2,1)$ generated by the Hermitian bilinears
\begin{align}
\label{GEN-DEF}
K_1=i(a_2a_3-a_2^{\dagger}a_{3}^{\dagger}),\quad K_2=i(a_{1}^{\dagger}a_{3}^{\dagger}-a_{1}a_{3}),\quad K_3=i(a_1a_{2}^{\dagger}-a_{1}^{\dagger}a_{2}),
\end{align}
satisfying the $\mathfrak{so}(2,1)$ commutation relations
\begin{align}
\label{Algebra}
[K_1,K_2]=-i K_3,\quad [K_2,K_3]=i K_1,\quad [K_3,K_1]=iK_2.
\end{align}
The reducible representations of the $SO(2,1)$ subgroup provided by the vectors \eqref{DEF-BASIS} will be considered. It will prove convenient to use the operators $b_i^{\dagger}$, $b_i$ defined by
\begin{align}
\label{b-Operators}
b_1=a_1,\qquad b_2=a_2,\qquad b_3=a_{3}^{\dagger}
\end{align}
instead of the standard creation/annihilation operators $a_i$, $a_i^{\dagger}$. On the basis \eqref{DEF-BASIS}, it is easily seen that one has the actions
\begin{subequations}
\label{Actions}
\begin{align}
&b_1\kket{m,n}{\beta}=\sqrt{m}\kket{m-1,n}{\beta+1}, &&b_1^{\dagger}\kket{m,n}{\beta}=\sqrt{m+1}\kket{m+1,n}{\beta-1},
\\
&b_2\kket{m,n}{\beta}=\sqrt{n}\kket{m,n-1}{\beta+1}, &&b_2^{\dagger}\kket{m,n}{\beta}=\sqrt{n+1}\kket{m,n+1}{\beta-1},
\\
&b_3\kket{m,n}{\beta}=\sqrt{m+n+\beta}\kket{m,n}{\beta+1}, &&b_3^{\dagger}\kket{m,n}{\beta}=\sqrt{m+n+\beta-1}\kket{m,n}{\beta-1}.
\end{align}
\end{subequations}
It is directly checked that the actions \eqref{Actions} define an infinite-dimensional representation of the Lie algebra \eqref{Algebra} on the oscillator states \eqref{DEF-BASIS}. The assertion that this representation is \emph{reducible} follows from the fact that the $\mathfrak{so}(2,1)$ Casimir operator $C=-K_1^2-K_2^2+K_3^2$ does not act as a multiple of the identity on \eqref{DEF-BASIS}.

We use the following notation. Let $\Lambda$ be an orthochronous transformation of $SO(2,1)$; this means that
\begin{align}
\label{DEF}
\Lambda^{t}\eta\Lambda=\eta,\qquad \Lambda_{33}\geqslant 1, 
\end{align}
where $A^{t}$ denotes the transpose matrix of $A$ and where $\eta=\mathrm{diag}(1,1,-1)$. Consider the unitary representation defined by
\begin{align}
\label{Unirep}
\mathcal{F}(\Lambda)=\exp\left(\sum_{i,k=1}^{3} B_{ik}b_{i}^{\dagger}b_{k}\right),
\end{align}
where $B_{ik}=-B_{ki}$. One has of course $\mathcal{F}(\Lambda)\mathcal{F}^{\dagger}(\Lambda)=1$. The transformations of the generators $b_{i}^{\dagger}$, $b_i$ under the action of $\mathcal{F}(\Lambda)$ are given by
\begin{align}
\label{Trans}
\mathcal{F}(\Lambda)b_i^{\dagger}\mathcal{F}^{\dagger}(\Lambda)=\sum_{k=1}^{3}\wt{\Lambda}_{ik}b_{k}^{\dagger},\qquad \mathcal{F}(\Lambda)b_{i}\mathcal{F}^{\dagger}(\Lambda)=\sum_{k=1}^{3}\wt{\Lambda}_{ik}b_{k},
\end{align}
where $\wt{\Lambda}=\eta\Lambda^{t}\eta$ stands for the inverse matrix of $\Lambda$: $\Lambda\wt{\Lambda}=1$. It is directly checked that $\mathcal{F}(\Lambda)$ satisfies
\begin{align}
\mathcal{F}(\Lambda\Delta)=\mathcal{F}(\Lambda)\mathcal{F}(\Delta),\qquad \Lambda,\Delta\in SO(2,1),
\end{align}
as should be for a group representation.
\section{The representation matrix elements as orthogonal polynomials}
In this section, it is shown that the matrix elements of the  $SO(2,1)$ unitary representation defined above are expressed in terms of orthogonal polynomials in two discrete variables.

The matrix elements of the unitary operator \eqref{Unirep} in the oscillator basis \eqref{DEF-BASIS} can be written as
\begin{align}
\label{Matrix-Elements}
\BBraket{\beta}{i,k}{\mathcal{F}(\Lambda)}{m,n}{\beta}=W_{i,k}^{(\beta)}\,M_{m,n}^{(\beta)}(i,k),
\end{align}
where $M_{0,0}^{(\beta)}(i,k)=1$ and where
\begin{align}
\label{Amplitude}
W_{i,k}^{(\beta)}=\BBraket{\beta}{i,k}{\mathcal{F}(\Lambda)}{0,0}{\beta}.
\end{align}
For notational ease, the explicit dependence of $\mathcal{F}(\Lambda)$ on $\Lambda$ will be omitted at times.
\subsection{Calculation of $W_{i,k}^{(\beta)}$}
To derive the explicit expression of the amplitude $W_{i,k}^{(\beta)}$, one first observes that
\begin{align*}
\BBraket{\beta+1}{i,k}{\mathcal{F}\,b_j}{0,0}{\beta}=0,
\end{align*}
for $j=1,2$. Since $\BBraket{\beta+1}{i,k}{\mathcal{F}b_j}{0,0}{\beta}=\BBraket{\beta}{i,k}{\mathcal{F}b_j\mathcal{F}^{\dagger}\mathcal{F}}{0,0}{\beta}$, one obtains, using \eqref{Trans}, the following system of difference equations for $W_{i,k}^{\beta}$:
\begin{subequations}
\label{Syst-1}
\begin{align}
\wt{\Lambda}_{11}\sqrt{i+1}\,W_{i+1,k}^{(\beta)}+\wt{\Lambda}_{12}\sqrt{k+1}\,W_{i,k+1}^{(\beta)}+\wt{\Lambda}_{13}\sqrt{i+k+\beta}\,W_{i,k}^{(\beta)}&=0,
\\
\wt{\Lambda}_{21}\sqrt{i+1}\,W_{i+1,k}^{(\beta)}+\wt{\Lambda}_{22}\sqrt{k+1}\,W_{i,k+1}^{(\beta)}+\wt{\Lambda}_{23}\sqrt{i+k+\beta}\,W_{i,k}^{(\beta)}&=0.
\end{align}
\end{subequations}
Using the fact that $\wt{\Lambda}\eta \wt{\Lambda}^{t}\eta=1$, it is readily seen that the solution to the system \eqref{Syst-1} is of the form
\begin{align*}
W_{i,k}^{(\beta)}=\sqrt{\frac{(\beta)_{i+k}}{i!k!}}\left(-\frac{\wt{\Lambda}_{31}}{\wt{\Lambda}_{33}}\right)^{i}\left(-\frac{\wt{\Lambda}_{32}}{\wt{\Lambda}_{33}}\right)^{k}W_{0,0}^{(\beta)},
\end{align*}
where $W_{0,0}^{(\beta)}=\BBraket{\beta}{0,0}{\mathcal{F}}{0,0}{\beta}$. The constant $W_{0,0}^{(\beta)}$ can be obtained from the normalization condition
\begin{align*}
1=\BBraket{\beta}{0,0}{\mathcal{F}^{\dagger}\mathcal{F}}{0,0}{\beta}=\sum_{i,k\geqslant 0}\BBraket{\beta}{i,k}{\mathcal{F}}{0,0}{\beta}\BBraket{\beta}{0,0}{\mathcal{F}^{\dagger}}{i,k}{\beta}=\sum_{i,k\geqslant 0}\rvert W_{i,k}^{(\beta)}\rvert^2.
\end{align*}
One can then use the formula
\begin{align*}
(1-z_1-z_2)^{-\beta}=\sum_{i,k}\frac{(\beta)_{i+k}}{i!k!}z_1^{i}z_2^{k},
\end{align*}
which holds provided that $|z_1|+|z_2|<1$. It is directly seen from \eqref{DEF} that this condition is identically satisfied and hence one finds that $W_{0,0}^{(\beta)}=[\wt{\Lambda}_{33}]^{-\beta}$. In terms of the matrix elements of $\Lambda$, the complete expression for the amplitude $W_{i,k}^{(\beta)}$ is thus found to be
\begin{align}
\label{Weight}
W_{i,k}^{(\beta)}=\sqrt{\frac{(\beta)_{i+k}}{i!k!}}\big(\Lambda_{33}\big)^{-\beta-i-k}\Lambda_{13}^{i}\Lambda_{23}^{k}.
\end{align}
\subsection{Raising relations}
To show that the $M_{m,n}^{(\beta)}(i,k)$ appearing in the matrix elements \eqref{Matrix-Elements} are polynomials of total degree $m+n$ in the two variables $i$, $k$, one can examine their raising relations, which are obtained as follows. One has on the one hand
\begin{align*}
&\BBraket{\beta}{i,k}{\mathcal{F}\,b_{1}^{\dagger}}{m,n}{\beta+1}=\sqrt{m+1}\,W_{i,k}^{(\beta)}\,M_{m+1,n}^{(\beta)}(i,k),
\\
&\BBraket{\beta}{i,k}{\mathcal{F}\,b_{2}^{\dagger}}{m,n}{\beta+1}=\sqrt{n+1}\,W_{i,k}^{(\beta)}\,M_{m,n+1}^{(\beta)}(i,k).
\end{align*}
On the other hand, using \eqref{Trans}, one has
\begin{align*}
&\BBraket{\beta}{i,k}{\mathcal{F}\,b_{1}^{\dagger}}{m,n}{\beta+1}=\BBraket{\beta}{i,k}{\mathcal{F}\,b_{1}^{\dagger}\,\mathcal{F}^{\dagger}\mathcal{F}}{m,n}{\beta+1}=\sum_{j=1}^{3}\wt{\Lambda}_{1j}\,\BBraket{\beta}{i,k}{b_{j}^{\dagger}\,\mathcal{F}}{m,n}{\beta+1},
\\
&\BBraket{\beta}{i,k}{\mathcal{F}\,b_{2}^{\dagger}}{m,n}{\beta+1}=\BBraket{\beta}{i,k}{\mathcal{F}\,b_{2}^{\dagger}\,\mathcal{F}^{\dagger}\mathcal{F}}{m,n}{\beta+1}=\sum_{j=1}^{3}\wt{\Lambda}_{2j}\,\BBraket{\beta}{i,k}{b_{j}^{\dagger}\,\mathcal{F}}{m,n}{\beta+1}.
\end{align*}
Upon combining the above relations, one obtains the raising relations
\begin{subequations}
\label{Raising}
\begin{align}
\begin{aligned}
\sqrt{\beta(m+1)}\,M_{m+1,n}^{(\beta)}(i,k)=&\frac{\Lambda_{11}}{\Lambda_{13}}\,i\,M_{m,n}^{(\beta+1)}(i-1,k)+\frac{\Lambda_{21}}{\Lambda_{23}}\,k\,M_{m,n}^{(\beta+1)}(i,k-1)
\\
&-\frac{\Lambda_{31}}{\Lambda_{33}}(i+k+\beta)\,M_{m,n}^{(\beta+1)}(i,k),
\end{aligned}
\\
\begin{aligned}
\sqrt{\beta(n+1)}\,M_{m,n+1}^{(\beta)}(i,k)=&\frac{\Lambda_{12}}{\Lambda_{13}}\,i\,M_{m,n}^{(\beta+1)}(i-1,k)+\frac{\Lambda_{22}}{\Lambda_{23}}\,k\,M_{m,n}^{(\beta+1)}(i,k-1)
\\
&-\frac{\Lambda_{32}}{\Lambda_{33}}(i+k+\beta)\,M_{m,n}^{(\beta+1)}(i,k).
\end{aligned}
\end{align}
\end{subequations}
By definition, one has $M_{-1,n}^{(\beta)}(i,k)=M_{m,-1}^{(\beta)}(i,k)=0$ and $M_{0,0}^{(\beta)}(i,k)=1$. Therefore, the formulas \eqref{Raising} can be used to construct $M_{m,n}^{(\beta)}(i,k)$ from $M_{0,0}^{(\beta)}(i,k)$ iteratively. Furthermore, it is observed that $M_{m,n}^{(\beta)}(i,k)$ are polynomials of total degree $m+n$ in the discrete variables $i$ and $k$.
\subsection{Orthogonality Relation}
It follows from the unitarity of the representation \eqref{Unirep} and the orthonormality of the oscillator states that the polynomials $M_{m,n}^{(\beta)}(i,k)$ are orthogonal with respect to the negative trinomial distribution. Indeed, one has
\begin{align*}
\BBraket{\beta}{m',n'}{\mathcal{F}^{\dagger}\mathcal{F}}{m,n}{\beta}=\sum_{i,k\geqslant0}\BBraket{\beta}{i,k}{\mathcal{F}}{m,n}{\beta}\BBraket{\beta}{m',n'}{\mathcal{F}^{\dagger}}{i,k}{\beta}=\delta_{mm'}\delta_{nn'}.
\end{align*}
Upon using \eqref{Matrix-Elements}, one finds that the polynomials satisfy the orthogonality relation
\begin{align}
\sum_{i,k\geqslant 0}w_{i,k}^{(\beta)}\,M_{m,n}^{(\beta)}(i,k)\,M_{m',n'}^{(\beta)}(i,k)=\delta_{mm'}\delta_{nn'},
\end{align}
where $w_{i,k}^{(\beta)}$ is the negative trinomial distribution
\begin{align}
\label{Neg-Tri}
w_{i,k}^{(\beta)}=\frac{(\beta)_{i+k}}{i!k!}(1-c_1-c_2)^{\beta}c_1^{i}c_2^{k},
\end{align}
with
\begin{align*}
c_1=\frac{\Lambda_{13}^2}{\Lambda_{33}^2},\qquad c_2=\frac{\Lambda_{23}^2}{\Lambda_{33}^2}.
\end{align*}
Recall that in view of \eqref{DEF}, the condition $|c_1|+|c_2|<1$ is identically satisfied for orthochronous elements of $SO(2,1)$.
\subsection{Lowering Relations}
Lowering relations for the polynomials $M_{m,n}^{(\beta)}(i,k)$ can also be obtained. To this end, one first considers the matrix elements
$\BBraket{\beta}{i,k}{\mathcal{F}\,b_{j}}{m,n}{\beta-1}$ with $j=1,2$. One has on the one hand
\begin{align*}
&\BBraket{\beta}{i,k}{\mathcal{F}\,b_{1}}{m,n}{\beta-1}=\sqrt{m}\,W_{i,k}^{(\beta)}\,M_{m-1,n}^{(\beta)}(i,k),
\\
&\BBraket{\beta}{i,k}{\mathcal{F}\,b_{2}}{m,n}{\beta-1}=\sqrt{n}\,W_{i,k}^{(\beta)}\,M_{m,n-1}^{(\beta)}(i,k),
\end{align*}
and on the other hand
\begin{align*}
\BBraket{\beta}{i,k}{\mathcal{F}\,b_{j}}{m,n}{\beta-1}=\BBraket{\beta}{i,k}{\mathcal{F}\,b_{j}\,\mathcal{F}^{\dagger}\mathcal{F}}{m,n}{\beta-1}=\sum_{\ell=1}^{3}\wt{\Lambda}_{j\ell}\,\BBraket{\beta}{i,k}{b_{\ell}\mathcal{F}}{m,n}{\beta-1}.
\end{align*}
Upon comparing the two expressions, a simple calculation yields
\begin{subequations}
\begin{align}
\begin{aligned}
\sqrt{\frac{m}{\beta-1}}\,M_{m-1,n}^{(\beta)}(i,k)=&\Lambda_{11}\Lambda_{13}\big[M_{m,n}^{(\beta-1)}(i+1,k)-M_{m,n}^{(\beta-1)}(i,k)\big]
\\
&+\Lambda_{21}\Lambda_{23}\big[M_{m,n}^{(\beta-1)}(i,k+1)-M_{m,n}^{(\beta-1)}(i,k)\big],
\end{aligned}
\\
\begin{aligned}
\sqrt{\frac{n}{\beta-1}}\,M_{m,n-1}^{(\beta)}(i,k)=&\Lambda_{12}\Lambda_{13}\big[M_{m,n}^{(\beta-1)}(i+1,k)-M_{m,n}^{(\beta-1)}(i,k)\big]
\\
&+\Lambda_{22}\Lambda_{23}\big[M_{m,n}^{(\beta-1)}(i,k+1)-M_{m,n}^{(\beta-1)}(i,k)\big].
\end{aligned}
\end{align}
\end{subequations}
\section{Duality}
In this section, a duality relation under the exchange of the degrees $m$, $n$ and the variables $i$, $k$ is derived for the multivariate polynomials $M_{m,n}^{(\beta)}(i,k)$. For the monic polynomials $R_{m,n}^{(\beta)}(i,k)$, which are obtained from the $M_{m,n}^{(\beta)}(i,k)$ by a normalization, this duality property is seen to take a particularly simple form.

The duality relation for the polynomials $M_{m,n}^{(\beta)}(i,k)$ is found by considering the matrix elements $\BBraket{\beta}{i,k}{\mathcal{F}^{\dagger}(\Lambda)}{m,n}{\beta}$ from two different points of view. First, one writes
\begin{align}
\label{1}
\BBraket{\beta}{i,k}{\mathcal{F}^{\dagger}(\Lambda)}{m,n}{\beta}=\wt{W}_{i,k}^{(\beta)}\wt{M}_{m,n}^{(\beta)}(i,k),
\end{align}
where $\wt{W}_{i,k}^{(\beta)}=\BBraket{\beta}{i,k}{\mathcal{F}^{\dagger}(\Lambda)}{0,0}{\beta}$ and $\wt{M}_{0,0}^{(\beta)}(i,k)=1$. On account of the fact that $\mathcal{F}^{\dagger}(\Lambda)=\mathcal{F}(\Lambda^{-1})=\mathcal{F}(\eta\Lambda^{t}\eta)$, it follows that
\begin{align*}
\wt{W}_{i,k}^{(\beta)}=\sqrt{\frac{(\beta)_{i+k}}{i!k!}}\big(\Lambda_{33}\big)^{-\beta-i-k}(-\Lambda_{31})^{i}(-\Lambda_{32})^{k},
\end{align*}
and that $\wt{M}_{m,n}^{(\beta)}(i,k)$ are the polynomials corresponding to the matrix $\wt{\Lambda}=\eta\Lambda^{t}\eta$. Second, one writes 
\begin{align}
\begin{aligned}
\label{2}
&\BBraket{\beta}{i,k}{\mathcal{F}^{\dagger}(\Lambda)}{m,n}{\beta}=\overline{\BBraket{\beta}{m,n}{\mathcal{F}(\Lambda)}{i,k}{\beta}}
\\
&=\BBraket{\beta}{m,n}{\mathcal{F}(\Lambda)}{i,k}{\beta}=W_{m,n}^{(\beta)}\,M_{i,k}^{(\beta)}(m,n),
\end{aligned}
\end{align}
where $\overline{x}$ stands for complex conjugation and where the reality of the matrix elements has been used. Upon comparing \eqref{1} and \eqref{2}, one obtains the duality relation
\begin{align}
\label{Duality-1}
M_{i,k}^{(\beta)}(m,n)=(-1)^{i+k}\sqrt{\frac{(\beta)_{i+k}}{i!k!}\frac{m!n!}{(\beta)_{m+n}}}\,\frac{\Lambda_{33}^{m+n}\Lambda_{31}^{i}\Lambda_{32}^{k}}{\Lambda_{33}^{i+k}\Lambda_{13}^{m}\Lambda_{23}^{n}}\;\wt{M}_{m,n}^{(\beta)}(i,k).
\end{align}
It is opportune here to introduce the monic polynomials $R_{m,n}^{(\beta)}(i,k)$ defined by
\begin{align*}
M_{m,n}^{(\beta)}(i,k)=(-1)^{m+n}\sqrt{\frac{(\beta)_{m+n}}{m!n!}}\frac{\Lambda_{31}^{m}\Lambda_{32}^{n}}{\Lambda_{33}^{m+n}}\,R_{m,n}^{(\beta)}(i,k).
\end{align*}
In terms of these polynomials, the duality relation \eqref{Duality-1} has the attractive expression
\begin{align}
\label{Duality-2}
R_{i,k}^{(\beta)}(m,n)=\wt{R}_{m,n}^{(\beta)}(i,k),
\end{align}
where the parameters appearing in the polynomials $\wt{R}_{m,n}^{(\beta)}(i,k)$ are those of the inverse matrix $\wt{\Lambda}=\eta\Lambda^{t}\eta$.
\section{Generating function and hypergeometric expression}
In this section, generating functions for the multivariate polynomials $M_{m,n}^{(\beta)}(i,k)$ and $R_{m,n}^{(\beta)}(i,k)$ are obtained using the group product. The generating function derived for the polynomials $R_{m,n}^{(\beta)}(i,k)$ is shown to coincide with the one defining the multivariate Meixner polynomials, which establishes that the polynomials $R_{m,n}^{(\beta)}(i,k)$ are precisely those defined in \cite{Griffiths-1975-06} and  \cite{Iliev-2012}. Using the results of \cite{Iliev-2012}, an explicit expression of the polynomials $R_{m,n}^{(\beta)}(i,k)$ in terms of Gel'fand-Aomoto hypergeometric series is given.

\subsection{Generating function}
Let $\Delta\in SO(2,1)$ be an arbitrary group element and consider the following generating function:
\begin{align}
\label{Gen}
G(\Delta)=\sum_{m,n\geqslant0}\sqrt{\frac{(\beta)_{m+n}}{m!n!}}\Delta_{33}^{-\beta-m-n}\Delta_{13}^{m}\Delta_{23}^{n}\,W_{i,k}^{(\beta)}\,M_{m,n}^{(\beta)}(i,k).
\end{align}
Given \eqref{Matrix-Elements}, one obviously has
\begin{align*}
G(\Delta)=\sum_{m,n\geqslant0}\sqrt{\frac{(\beta)_{m+n}}{m!n!}}\Delta_{33}^{-\beta-m-n}\Delta_{13}^{m}\Delta_{23}^{n}\;\BBraket{\beta}{i,k}{\mathcal{F}(\Lambda)}{m,n}{\beta}.
\end{align*}
In view of \eqref{Amplitude} and \eqref{Weight}, one finds that the above expression for $G(\Delta)$ can be written in the form
\begin{align*}
G(\Delta)=\BBraket{\beta}{i,k}{\mathcal{F}(\Lambda)\mathcal{F}(\Delta)}{0,0}{\beta}=\BBraket{\beta}{i,k}{\mathcal{F}(\Lambda\Delta)}{0,0}{\beta}.
\end{align*}
Upon using again the expression \eqref{Weight}, one arrives at
\begin{align}
\label{4}
G(\Delta)=\sqrt{\frac{(\beta)_{i+k}}{i!k!}}[(\Lambda\Delta)_{33}]^{-\beta-i-k}[(\Lambda\Delta)_{13}]^{i}[(\Lambda\Delta)_{23}]^{k},
\end{align}
where $(\Lambda\Delta)_{ij}$ are the matrix elements of the matrix $\Lambda\cdot \Delta$. Comparing \eqref{Gen} with \eqref{4} using the expression \eqref{Weight}, one arrives at the expression
\begin{align*}
&\left(\Delta_{33}+\frac{\Delta_{13}\Lambda_{11}}{\Lambda_{13}}+\frac{\Delta_{23}\Lambda_{12}}{\Lambda_{13}}\right)^{i}\left(\Delta_{33}+\frac{\Delta_{13}\Lambda_{21}}{\Lambda_{23}}+\frac{\Delta_{23}\Lambda_{22}}{\Lambda_{23}}\right)^{k}
\\
&\times\left(\Delta_{33}+\frac{\Delta_{13}\Lambda_{31}}{\Lambda_{33}}+\frac{\Delta_{23}\Lambda_{32}}{\Lambda_{33}}\right)^{-\beta-i-k}
=\sum_{m,n\geqslant 0}\sqrt{\frac{(\beta)_{m+n}}{m!n!}}(\Delta_{33})^{-\beta-m-n}\Delta_{13}^{m}\Delta_{23}^{n}\,M_{m,n}^{(\beta)}(i,k).
\end{align*}
Since $\Delta$ is arbitrary, one can choose the parametrization
\begin{align*}
\Delta_{13}=\delta_1,\qquad \Delta_{23}=\delta_2,\quad \Delta_{33}=1,
\end{align*}
which gives the generating function for the polynomials $M_{m,n}^{(\beta)}(i,k)$
\begin{align}
\begin{aligned}
&\left(1+\frac{\Lambda_{11}}{\Lambda_{13}}\delta_1+\frac{\Lambda_{12}}{\Lambda_{13}}\delta_2\right)^{i}\left(1+\frac{\Lambda_{21}}{\Lambda_{23}}\delta_1+\frac{\Lambda_{22}}{\Lambda_{23}}\delta_2\right)^{k}
\\
&\times\left(1+\frac{\Lambda_{31}}{\Lambda_{33}}\delta_1+\frac{\Lambda_{32}}{\Lambda_{33}}\delta_2\right)^{-\beta-i-k}
=\sum_{m,n\geqslant 0}\sqrt{\frac{(\beta)_{m+n}}{m!n!}}\,M_{m,n}^{(\beta)}(i,k)\,\delta_1^{m}\delta_2^{n}.
\end{aligned}
\end{align}
Upon choosing instead the parametrization
\begin{align*}
\Delta_{13}=-\frac{\Lambda_{33}}{\Lambda_{31}}z_1,\quad \Delta_{23}=-\frac{\Lambda_{33}}{\Lambda_{32}}z_2,\quad \Delta_{33}=1,
\end{align*}
one finds the following generating function for the monic polynomials $R_{m,n}^{(\beta)}(i,k)$:
\begin{align}
\label{Generating}
\begin{aligned}
&(1-z_1-z_2)^{-\beta-i-k}(1-u_{11}z_1-u_{12}z_2)^{i}(1-u_{21}z_1-u_{22}z_2)^{k}
\\
&=\sum_{m,n\geqslant 0}\frac{(\beta)_{m+n}}{m!n!}\,R_{m,n}^{(\beta)}(i,k)\,z_1^{m}z_{2}^{n},
\end{aligned}
\end{align}
where
\begin{align}
\label{U-Parameters}
u_{11}=\frac{\Lambda_{11}\Lambda_{33}}{\Lambda_{13}\Lambda_{31}}, \qquad u_{12}=\frac{\Lambda_{12}\Lambda_{33}}{\Lambda_{13}\Lambda_{32}},\qquad  u_{21}=\frac{\Lambda_{21}\Lambda_{33}}{\Lambda_{23}\Lambda_{31}}, \qquad u_{22}=\frac{\Lambda_{22}\Lambda_{33}}{\Lambda_{23}\Lambda_{32}}.
\end{align}
The generating function \eqref{Generating} is identical to the one considered in \cite{Griffiths-1975-06,Iliev-2012} which is taken to define the bivariate Meixner polynomials. Thus it follows that the monic polynomials $R_{m,n}^{(\beta)}(i,k)$ are the bivariate Meixner polynomials. Note that here $\beta>1$ is an integer, but it is directly seen that the polynomials $R_{m,n}^{(\beta)}(i,k)$ can be defined for $\beta\in \mathbb{R}$.
\subsection{Hypergeometric expression}
In \cite{Iliev-2012}, Iliev obtained an explicit expression for the Meixner polynomials $R_{m,n}^{(\beta)}(i,k)$ in terms of a Gel'fand-Aomoto hypergeometric series. Using his result, one writes
\begin{align}
\label{Hypergeometric}
\begin{aligned}
&R_{m,n}^{(\beta)}(i,k)=\sum_{\mu,\nu,\rho,\sigma\geqslant 0}\frac{(-m)_{\mu+\nu}(-n)_{\rho+\sigma}(-i)_{\mu+\rho}(-k)_{\nu+\sigma}}{\mu!\nu!\rho!\sigma!(\beta)_{\mu+\nu+\rho+\sigma}}
\\
&\times(1-u_{11})^{\mu}(1-u_{21})^{\nu}(1-u_{12})^{\rho}(1-u_{22})^{\sigma},
\end{aligned}
\end{align}
where the parameters $u_{ij}$ are given by \eqref{U-Parameters}. Since $(-m)_{k}=0$ for $k>m$, it is clear that the summation in \eqref{Hypergeometric} is finite.
\section{Recurrence relations and difference equations}
In this section, the recurrence relations and the difference equations satisfied by the bivariate Meixner polynomials $M_{m,n}^{(\beta)}(i,k)$ and $R_{m,n}^{(\beta)}(i,k)$ are derived. These relations have been obtained by Iliev in \cite{Iliev-2012}. It is however interesting to see how easily these relations follow from the group-theoretical interpretation.
\subsection{Recurrence relations}
To obtain the recurrence relations satisfied by the Meixner polynomials, one considers the matrix elements $\BBraket{\beta}{i,k}{b_{j}^{\dagger}b_{j}\mathcal{F}}{m,n}{\beta}$ for $j=1,2$. On the one hand one has
\begin{align}
\label{Dompe}
\begin{aligned}
&\BBraket{\beta}{i,k}{b_{1}^{\dagger}b_1\,\mathcal{F}}{m,n}{\beta}=i\;\BBraket{\beta}{i,k}{\mathcal{F}}{m,n}{\beta},
\\
&\BBraket{\beta}{i,k}{b_{2}^{\dagger}b_2\,\mathcal{F}}{m,n}{\beta}=k\;\BBraket{\beta}{i,k}{\mathcal{F}}{m,n}{\beta},
\end{aligned}
\end{align}
and on the other hand
\begin{align}
\label{Dompe-2}
\begin{aligned}
&\BBraket{\beta}{i,k}{b_{j}^{\dagger}b_j\,\mathcal{F}}{m,n}{\beta}=\BBraket{\beta}{i,k}{\mathcal{F}\mathcal{F}^{\dagger}b_{j}^{\dagger}b_{j}\mathcal{F}}{m,n}{\beta}
\\
&\sum_{\ell=1}^{3}\sum_{r=1}^{3}\Lambda_{j\ell}\Lambda_{jr}\;\BBraket{\beta}{i,k}{\mathcal{F}\,b_{\ell}^{\dagger}b_{r}}{m,n}{\beta}.
\end{aligned}
\end{align}
Upon comparing \eqref{Dompe} with \eqref{Dompe-2}, one directly obtains
\begin{subequations}
\label{Rec-1}
\begin{align}
&i\,M_{m,n}^{(\beta)}(i,k)=\Big[m\Lambda_{11}^2+n \Lambda_{12}^2+(m+n+\beta)\Lambda_{13}^2\Big]M_{m,n}^{(\beta)}(i,k)\nonumber
\\
&+\Lambda_{11}\Lambda_{12}\Big[\sqrt{m(n+1)}M_{m-1,n+1}^{(\beta)}(i,k)+\sqrt{n(m+1)}M_{m+1,n-1}^{(\beta)}(i,k)\Big]
\\
&+\Lambda_{11}\Lambda_{13}\Big[\sqrt{m(m+n+\beta-1)}M_{m-1,n}^{(\beta)}(i,k)+\sqrt{(m+1)(m+n+\beta)}M_{m+1,n}^{(\beta)}(i,k)\Big]\nonumber
\\
&+\Lambda_{12}\Lambda_{13}\Big[\sqrt{n(m+n+\beta-1)}M_{m,n-1}^{(\beta)}(i,k)+\sqrt{(n+1)(n+m+\beta)}M_{m,n+1}^{(\beta)}(i,k)\Big]\nonumber
\end{align}
\begin{align}
&k\,M_{m,n}^{(\beta)}(i,k)=\Big[m\Lambda_{21}^2+n \Lambda_{22}^2+(m+n+\beta)\Lambda_{23}^2\Big]M_{m,n}^{(\beta)}(i,k)\nonumber
\\
&+\Lambda_{21}\Lambda_{22}\Big[\sqrt{m(n+1)}M_{m-1,n+1}^{(\beta)}(i,k)+\sqrt{n(m+1)}M_{m+1,n-1}^{(\beta)}(i,k)\Big]
\\
&+\Lambda_{21}\Lambda_{23}\Big[\sqrt{m(m+n+\beta-1)}M_{m-1,n}^{(\beta)}(i,k)+\sqrt{(m+1)(m+n+\beta)}M_{m+1,n}^{(\beta)}(i,k)\Big]\nonumber
\\
&+\Lambda_{22}\Lambda_{23}\Big[\sqrt{n(m+n+\beta-1)}M_{m,n-1}^{(\beta)}(i,k)+\sqrt{(n+1)(n+m+\beta)}M_{m,n+1}^{(\beta)}(i,k)\Big]\nonumber
\end{align}
\end{subequations}
\normalsize
For the monic Meixner polynomials $R_{m,n}^{(\beta)}(i,k)$, one finds from \eqref{Rec-1}
\begin{subequations}
\label{Rec-2}
\begin{align}
i\,R_{m,n}^{(\beta)}(i,k)&=\Big[m\Lambda_{11}^2+n \Lambda_{12}^2+(m+n+\beta)\Lambda_{13}^2\Big]R_{m,n}^{(\beta)}(i,k)\nonumber
\\
\label{Rec-2-a}
&+\frac{\Lambda_{11}\Lambda_{12}\Lambda_{32}}{\Lambda_{31}}\,m\,R_{m-1,n+1}^{(\beta)}(i,k)+\frac{\Lambda_{11}\Lambda_{12}\Lambda_{31}}{\Lambda_{32}}\,n\,R_{m+1,n-1}^{(\beta)}(i,k)
\\
&-\frac{\Lambda_{11}\Lambda_{13}\Lambda_{33}}{\Lambda_{31}}\,m\,R_{m-1,n}^{(\beta)}(i,k)-\frac{\Lambda_{11}\Lambda_{13}\Lambda_{31}}{\Lambda_{33}}\,(m+n+\beta)\,R_{m+1,n}^{(\beta)}(i,k)\nonumber
\\
&-\frac{\Lambda_{12}\Lambda_{13}\Lambda_{33}}{\Lambda_{32}}\,n\,R_{m,n-1}^{(\beta)}(i,k) -\frac{\Lambda_{12}\Lambda_{13}\Lambda_{32}}{\Lambda_{33}}\,(m+n+\beta)\,R_{m,n+1}^{(\beta)}(i,k),\nonumber
\\
k\,R_{m,n}^{(\beta)}(i,k)&=\Big[m\Lambda_{21}^2+n \Lambda_{22}^2+(m+n+\beta)\Lambda_{23}^2\Big]R_{m,n}^{(\beta)}(i,k)\nonumber
\\
&+\frac{\Lambda_{21}\Lambda_{22}\Lambda_{32}}{\Lambda_{31}}\,m\,R_{m-1,n+1}^{(\beta)}(i,k)+\frac{\Lambda_{21}\Lambda_{22}\Lambda_{31}}{\Lambda_{32}}\,n\,R_{m+1,n-1}^{(\beta)}(i,k)
\\
&-\frac{\Lambda_{21}\Lambda_{23}\Lambda_{33}}{\Lambda_{31}}\,m\,R_{m-1,n}^{(\beta)}(i,k)-\frac{\Lambda_{21}\Lambda_{23}\Lambda_{31}}{\Lambda_{33}}\,(m+n+\beta)\,R_{m+1,n}^{(\beta)}(i,k)\nonumber
\\
&-\frac{\Lambda_{22}\Lambda_{23}\Lambda_{33}}{\Lambda_{32}}\,n\,R_{m,n-1}^{(\beta)}(i,k) -\frac{\Lambda_{22}\Lambda_{23}\Lambda_{32}}{\Lambda_{33}}\,(m+n+\beta)\,R_{m,n+1}^{(\beta)}(i,k).\nonumber
\end{align}
\end{subequations}
\subsection{Difference equations}
To obtain the difference equations satisfied by the polynomials $R_{m,n}^{(\beta)}(i,k)$, one could consider the matrix elements $\BBraket{\beta}{i,k}{\mathcal{F}b_{j}^{\dagger}b_{j}}{m,n}{\beta}$ for $j=1,2$ and proceed along the same lines as for the recurrence relations. It is however more elegant to proceed directly from the recurrence relations \eqref{Rec-2} and to use the duality property \eqref{Duality-2} of the monic bivariate Meixner polynomials. To illustrate the method, consider the left-hand side of \eqref{Rec-2-a}. Upon using the duality \eqref{Duality-2}, one may write
\begin{align*}
i\,R_{m,n}^{(\beta)}(i,k)=i\,\wt{R}_{i,k}^{(\beta)}(m,n)\rightarrow m\,\wt{R}^{(\beta)}_{m,n}(i,k),
\end{align*}
where in the last step the replacements $m\leftrightarrow i$, $n\leftrightarrow k$ were performed. Since $\wt{R}_{m,n}^{(\beta)}(i,k)$, is obtained from $R_{m,n}^{(\beta)}(i,k)$ by replacing the parameters of $\Lambda$ by the parameters of the inverse matrix $\wt{\Lambda}=\eta \Lambda^{t}\eta$, it is seen that the recurrence relations \eqref{Rec-2} can be converted into difference equations by taking $m\leftrightarrow i$, $n\leftrightarrow k$ and replacing the parameters of $\Lambda$ by those of the inverse. This yields
\begin{subequations}
\label{Ultra-2}
\begin{align}
m\,R_{m,n}^{(\beta)}(i,k)&=\Big[i\,\Lambda_{11}^2+k\,\Lambda_{21}^2+(i+k+\beta)\,\Lambda_{31}^2\Big]R_{m,n}^{(\beta)}(i,k)\nonumber
\\
&+\frac{\Lambda_{11}\Lambda_{21}\Lambda_{23}}{\Lambda_{13}}\,i\,R_{m,n}^{(\beta)}(i-1,k+1)+\frac{\Lambda_{11}\Lambda_{21}\Lambda_{13}}{\Lambda_{23}}\,k\,R_{m,n}^{(\beta)}(i+1,k-1)
\\
&-\frac{\Lambda_{11}\Lambda_{31}\Lambda_{33}}{\Lambda_{13}}\,i\,R_{m,n}^{(\beta)}(i-1,k)-\frac{\Lambda_{11}\Lambda_{31}\Lambda_{13}}{\Lambda_{33}}\,(i+k+\beta)\,R_{m,n}^{(\beta)}(i+1,k)\nonumber
\\
&-\frac{\Lambda_{21}\Lambda_{31}\Lambda_{33}}{\Lambda_{23}}\,k\,R_{m,n}^{(\beta)}(i,k-1) -\frac{\Lambda_{21}\Lambda_{31}\Lambda_{23}}{\Lambda_{33}}\,(i+k+\beta)\,R_{m,n}^{(\beta)}(i,k+1),\nonumber
\end{align}
\begin{align}
n\,R_{m,n}^{(\beta)}(i,k)&=\Big[i\,\Lambda_{12}^2+k\,\Lambda_{22}^2+(i+k+\beta)\,\Lambda_{32}^2\Big]R_{m,n}^{(\beta)}(i,k)\nonumber
\\
&+\frac{\Lambda_{12}\Lambda_{22}\Lambda_{23}}{\Lambda_{13}}\,i\,R_{m,n}^{(\beta)}(i-1,k+1)+\frac{\Lambda_{12}\Lambda_{22}\Lambda_{13}}{\Lambda_{23}}\,k\,R_{m,n}^{(\beta)}(i+1,k-1)
\\
&-\frac{\Lambda_{12}\Lambda_{32}\Lambda_{33}}{\Lambda_{13}}\,i\,R_{m,n}^{(\beta)}(i-1,k)-\frac{\Lambda_{12}\Lambda_{32}\Lambda_{13}}{\Lambda_{33}}\,(i+k+\beta)\,R_{m,n}^{(\beta)}(i+1,k)\nonumber
\\
&-\frac{\Lambda_{22}\Lambda_{32}\Lambda_{33}}{\Lambda_{23}}\,k\,R_{m,n}^{(\beta)}(i,k-1) -\frac{\Lambda_{22}\Lambda_{32}\Lambda_{23}}{\Lambda_{33}}\,(i+k+\beta)\,R_{m,n}^{(\beta)}(i,k+1)\nonumber.
\end{align}
\end{subequations}
The same method can be applied to obtain the difference equations satisfied by the polynomials $M_{m,n}^{(\beta)}(i,k)$. Note that the difference equations \eqref{Ultra-2} can be combined to give the following nearest neighbour difference equation for the polynomials $R_{m,n}^{(\beta)}(i,k)$:
\begin{align*}
&\left[\frac{m}{\Lambda_{11}\Lambda_{21}}-\frac{n}{\Lambda_{12}\Lambda_{22}}\right]R_{m,n}^{(\beta)}(i,k)=\\
&\left[i\left(\frac{\Lambda_{11}}{\Lambda_{21}}-\frac{\Lambda_{12}}{\Lambda_{22}}\right)+k\left(\frac{\Lambda_{21}}{\Lambda_{11}}-\frac{\Lambda_{22}}{\Lambda_{12}}\right)+(i+k+\beta)\left(\frac{\Lambda_{31}^2}{\Lambda_{11}\Lambda_{21}}-\frac{\Lambda_{32}^2}{\Lambda_{12}\Lambda_{22}}\right)\right]R_{m,n}^{(\beta)}(i,k)
\\
&+i\left[\frac{\Lambda_{32}\Lambda_{33}}{\Lambda_{13}\Lambda_{22}}-\frac{\Lambda_{31}\Lambda_{33}}{\Lambda_{21}\Lambda_{13}}\right]R_{m,n}^{(\beta)}(i-1,k)+(i+k+\beta)\left[\frac{\Lambda_{13}\Lambda_{32}}{\Lambda_{22}\Lambda_{33}}-\frac{\Lambda_{13}\Lambda_{31}}{\Lambda_{21}\Lambda_{33}}\right]R_{m,n}^{(\beta)}(i+1,k)
\\
&+k\left[\frac{\Lambda_{32}\Lambda_{33}}{\Lambda_{12}\Lambda_{23}}-\frac{\Lambda_{31}\Lambda_{33}}{\Lambda_{11}\Lambda_{23}}\right]R_{m,n}^{(\beta)}(i,k-1)+(i+k+\beta)\left[\frac{\Lambda_{23}\Lambda_{32}}{\Lambda_{12}\Lambda_{33}}-\frac{\Lambda_{23}\Lambda_{31}}{\Lambda_{11}\Lambda_{33}}\right]R_{m,n}^{(\beta)}(i,k+1).
\end{align*}
\normalsize
A similar formulas holds for the bivariate Krawtchouk polynomials \cite{Grunbaum-2007-05}.
\section{One-parameter subgroups and univariate Meixner \& Krawtchouk polynomials}
It has been assumed so far that the entries $\Lambda_{ij}$ of the $SO(2,1)$ parameter matrix for the bivariate Meixner polynomials are non-zero. In this section, degenerate cases corresponding to natural one-parameter subgroups of $SO(2,1)$ shall be considered. In particular, it will be shown that for transformations $\Lambda$ belonging to hyperbolic subgroups, the matrix elements $\BBraket{\beta}{i,k}{\mathcal{F}(\Lambda)}{m,n}{\beta}$ are given in terms of the standard (univariate) Meixner polynomials and that for transformations $\Lambda$ belonging to the elliptic subgroup, the matrix elements  $\BBraket{\beta}{i,k}{\mathcal{F}(\Lambda)}{m,n}{\beta}$ are expressed in terms of standard (univariate) Krawtchouk polynomials. The one-variable Meixner polynomials $M_{n}(x;\delta;c)$ are defined by their explicit expression \cite{Koekoek-2010}
\begin{align}
\label{Meixner}
M_{n}(x;\delta;c)=\pFq{2}{1}{-n,-x}{\delta}{1-\frac{1}{c}},
\end{align}
where ${}_{2}F_{1}$ is the Gauss hypergeometric function. The monic Meixner polynomials $m_{n}(x)$ defined through $M_{n}(x;\delta;c)=\frac{1}{(\delta)_{n}}\left(\frac{c-1}{c}\right)^{n}m_{n}(x)$ obey the three term recurrence relation
\begin{align}
\label{Recu-Meix}
x\,m_{n}(x)=m_{n+1}(x)+\frac{n+(n+\delta)c}{1-c}\,m_{n}(x)+\frac{n(n+\delta-1)c}{(1-c)^2}\,m_{n-1}(x),
\end{align}
with $m_{-1}(x)=0$ and $m_0(x)=1$. The one-variable Krawtchouk polynomials are denoted $K_{n}(x;p;N)$ and have the expression
\begin{align}
\label{Krawtchouk}
K_{n}(x;p;N)=\pFq{2}{1}{-n,-x}{-N}{\frac{1}{p}},
\end{align}
where $N$ is a positive integer.
\subsection{Hyperbolic subgroups: Meixner polynomials}
Consider the two hyperbolic one-parameter subgroups of $SO(2,1)$ which have as representative elements the following matrices:
\begin{subequations}
\begin{align}
\label{Mat-a}
\Xi(\xi)&=
\begin{pmatrix}
\cosh \xi & 0 & \sinh\xi
\\
0 & 1 & 0
\\
\sinh \xi & 0 & \cosh \xi
\end{pmatrix},
\\
\Psi(\psi)&=
\begin{pmatrix}
1 & 0 & 0
\\
0 & \cosh \psi & \sinh \psi
\\
0 & \sinh \psi & \cosh \psi
\end{pmatrix}.
\end{align}
\end{subequations}
The pseudo-rotations $\Xi(\xi)$ and $\Psi(\psi)$ are unitarily represented by the operators
\begin{align*}
\mathcal{F}(\Xi(\xi))=e^{-i\xi K_2},\qquad \mathcal{F}(\Psi(\psi))=e^{-i\psi K_1},
\end{align*}
where $K_1$, $K_2$ are given by \eqref{GEN-DEF}. The matrix elements $\BBraket{\beta}{i,k}{\mathcal{F}(\Xi(\xi))}{m,n}{\beta}$ and $\BBraket{\beta}{i,k}{\mathcal{F}(\Psi(\psi))}{m,n}{\beta}$ can be evaluated using the approach of Section 3. Let us give the details of the calculation of the matrix elements of $\mathcal{F}(\Xi(\xi))$. Since $\Xi(\xi)$ leaves $b_{2}$ and $b_2^{\dagger}$ unaffected and hence acts in a trivial way on the second quantum number, it is readily seen that
\begin{align*}
\BBraket{\beta}{i,k}{\mathcal{F}(\Xi)}{m,n}{\beta}=\delta_{kn}\;\BBraket{\beta}{i}{\mathcal{F}(\Xi)}{m}{\beta},
\end{align*}
where the dependence on $\xi$ and on the second quantum number have been suppressed for notational convenience. Given that
\begin{align*}
\mathcal{F}^{\dagger}(\Xi)b_1\mathcal{F}(\Xi)=b_1\cosh\xi+b_3\sinh\xi,\quad \mathcal{F}^{\dagger}(\Xi)b_1^{\dagger}\mathcal{F}(\Xi)=b_1^{\dagger}\cosh\xi+b_3^{\dagger}\sinh \xi,
\end{align*}
the identity $\BBraket{\beta}{i}{b_{1}^{\dagger}b_1\mathcal{F}(\Xi)}{m}{\beta}=\BBraket{\beta}{i}{\mathcal{F}(\Xi)\mathcal{F}^{\dagger}(\Xi)b_{1}^{\dagger}b_1\,\mathcal{F}(\Xi)}{m}{\beta}$ yields the recurrence relation
\begin{align*}
i\;\BBraket{\beta}{i}{\mathcal{F}(\Xi)}{m}{\beta}&=\big[m\cosh^2\xi+(m+\gamma)\sinh^2\xi\big]\,\BBraket{\beta}{i}{\mathcal{F}(\Xi)}{m}{\beta}
\\
&+\cosh\xi\sinh\xi\sqrt{m(m+\gamma-1)}\,\BBraket{\beta}{i}{\mathcal{F}(\Xi)}{m-1}{\beta}
\\
&+\cosh\xi\sinh\xi\sqrt{(m+1)(m+\gamma)}\,\BBraket{\beta}{i}{\mathcal{F}(\Xi)}{m+1}{\beta},
\end{align*}
where $\gamma=n+\beta$. Upon taking 
\begin{align*}
\BBraket{\beta}{i}{\mathcal{F}(\Xi)}{m}{\beta}=\BBraket{\beta}{i}{\mathcal{F}(\Xi)}{0}{\beta}\,\sqrt{\frac{1}{m!(\gamma)_{m}}}\,(\cosh \xi\sinh\xi)^{-n}\,P_{m}(i),
\end{align*}
where $P_{0}(i)=1$, it is seen that $P_{m}(i)$ satisfies the three-term recurrence relation \eqref{Recu-Meix} of the monic Meixner polynomials with $c=\tanh^{2}\xi$ and $\delta=\gamma=\beta+n$. One thus has
\begin{align*}
\BBraket{\beta}{i}{\mathcal{F}(\Xi)}{m}{\beta}=\BBraket{\beta}{i}{\mathcal{F}(\Xi)}{0}{\beta}\;(-1)^{m}\sqrt{\frac{(n+\beta)_{m}}{m!}}\tanh^{m}\xi\,M_{m}(i;\beta+n;\tanh^2\xi),
\end{align*}
where $M_{n}(x;\delta;c)$ are the univariate Meixner polynomials. There remains to evaluate the amplitude $\BBraket{\beta}{i}{\mathcal{F}(\Xi)}{0}{\beta}$. This can be done using the identity $\BBraket{\beta+1}{i}{\mathcal{F}(\Xi)b_1}{0}{\beta}=0$ which gives the two-term recurrence relation
\begin{align*}
\BBraket{\beta}{i+1}{\mathcal{F}(\Xi)}{0}{\beta}=\tanh\xi\,\sqrt{\frac{i+\gamma}{i+1}}\;\BBraket{\beta}{i}{\mathcal{F}(\Xi)}{0}{\beta},
\end{align*}
that has for solution
\begin{align*}
\BBraket{\beta}{i}{\mathcal{F}(\Xi)}{0}{\beta}=\tanh^{i}\xi\sqrt{\frac{(\gamma)_i}{i!}}\,\BBraket{\beta}{0}{\mathcal{F}(\Xi)}{0}{\beta}.
\end{align*}
Since one has
\begin{align*}
1=\bbraket{\beta}{0}{0}{\beta}=\sum_{i\geqslant 0}\BBraket{\beta}{0}{\mathcal{F}^{\dagger}(\Xi)}{i}{\beta}\,\BBraket{\beta}{i}{\mathcal{F}(\Xi)}{0}{\beta}=\sum_{i\geqslant 0}\frac{(\gamma)_i}{i!}\tanh^{2i}\xi \;\rvert\BBraket{\beta}{0}{\mathcal{F}(\Xi)}{0}{\beta}\rvert^{2},
\end{align*}
it follows that
\begin{align*}
\BBraket{\beta}{i}{\mathcal{F}(\Xi)}{0}{\beta}=\sqrt{\frac{(\gamma)_{i}}{i!}}\,\cosh^{-\gamma-i}\xi\sinh^{i}\xi.
\end{align*}
The matrix elements of the one-parameter hyperbolic elements $\Xi(\xi)$ are thus given by
\begin{align}
\begin{aligned}
\label{First}
&\BBraket{\beta}{i,k}{\mathcal{F}(\Xi(\xi))}{m,n}{\beta}\\
&=\delta_{kn}\,(-1)^{m}\sqrt{\frac{(k+\beta)_{i}(k+\beta)_{m}}{i!m!}}\,\cosh^{-k-\beta}\xi\,\tanh^{i+m}\xi\,M_{m}(i;k+\beta;\tanh^2\xi).
\end{aligned}
\end{align}
In a similar fashion, one obtains for the matrix elements of $\mathcal{F}(\Psi)$
\begin{align}
\begin{aligned}
\label{Second}
&\BBraket{\beta}{i,k}{\mathcal{F}(\Psi(\psi))}{m,n}{\beta}\\
&=\delta_{im}\,(-1)^{n}\sqrt{\frac{(i+\beta)_{k}(i+\beta)_{n}}{k!n!}}\,\cosh^{-i-\beta}\psi\,\tanh^{k+n}\psi\,M_{n}(k;i+\beta;\tanh^2\psi).
\end{aligned}
\end{align}
\subsection{Elliptic subgroup: Krawtchouk polynomials}
The group $SO(2,1)$ also has a one-parameter elliptic subgroup which has for representative element the matrix
\begin{align}
\label{Rotation}
R(\theta)=
\begin{pmatrix}
\cos \theta & \sin \theta & 0
\\
-\sin\theta & \cos \theta & 0
\\
0 & 0 & 1
\end{pmatrix},
\end{align}
which is unitarily represented by $\mathcal{F}(R(\theta))=e^{i\theta K_3}$. The matrix elements $\BBraket{\beta}{i,k}{\mathcal{F}(R(\theta))}{m,n}{\beta}$ can be evaluated using the same approach as the one adopted above. Since the details of similar computations are found in \cite{Genest-2013-06}, we only give the result which reads
\begin{align}
\label{Ultra}
\begin{aligned}
&\BBraket{\beta}{i,k}{\mathcal{F}(R(\theta))}{m,n}{\beta}
\\
&=\delta_{i+k,m+n}\,(-1)^{k}\sqrt{\binom{i+k}{k}\binom{i+k}{n}}\,\cos^{i+k}\theta\tan^{k+n}\theta\,K_{n}(k;\sin^{2}\theta;i+k),
\end{aligned}
\end{align}
where $K_{n}(x;p;N)$ is given by \eqref{Krawtchouk} and where $\binom{N}{i}$ stands for the binomial coefficient. Note that the formula \eqref{Ultra} does not define proper Krawtchouk polynomials since here $N$ is not an independent parameter as the variable $i+k$ occurs in its place.
\section{Addition formulas}
In this section, the group product is used to derive a general addition formula for the bivariate Meixner polynomials.
\subsection{General addition formula}
Let $A$, $B$ and $C$ be $SO(2,1)$ elements such that $C=A\cdot B$ with unitary representations $\mathcal{F}(A)$, $\mathcal{F}(B)$ and $\mathcal{F}(C)$. For a given value of $\beta$, to each of these elements is associated a system of bivariate Meixner polynomials denoted by $M_{m,n}^{(\beta)}(i,k;A)$, $M_{m,n}^{(\beta)}(i,k;B)$ and $M_{m,n}^{(\beta)}(i,k;C)$. Since $\mathcal{F}(C)=\mathcal{F}(A)\mathcal{F}(B)$, it follows that
\begin{align}
\label{Decompo}
\BBraket{\beta}{i,k}{\mathcal{F}(C)}{m,n}{\beta}=\sum_{\rho,\sigma\geqslant0}\BBraket{\beta}{i,k}{\mathcal{F}(A)}{\rho,\sigma}{\beta}\,\BBraket{\beta}{\rho,\sigma}{\mathcal{F}(B)}{m,n}{\beta}.
\end{align}
In terms of the polynomials $M_{m,n}^{(\beta)}(i,k)$, this identity translates into the addition formula
\begin{align}
\label{Addition}
\left(\frac{W_{i,k}^{(\beta)}(C)}{W_{i,k}^{(\beta)}(A)}\right)\,M_{m,n}^{(\beta)}(i,k;C)=\sum_{\rho,\sigma\geqslant 0}W_{\rho,\sigma}^{(\beta)}(B)M_{\rho,\sigma}^{(\beta)}(i,k;A)\,M_{m,n}^{(\beta)}(\rho,\sigma;B).
\end{align}
\subsection{Special case I: product of two hyperbolic elements}
Consider the case where the $SO(2,1)$ parameter matrix for the bivariate Meixner polynomials $R_{m,n}^{(\beta)}(i,k)$ is of the form
\small
\begin{align}
\label{Matrix}
\Lambda=\Psi(\psi)\cdot \Xi(\xi)=
\begin{pmatrix}
\cosh \xi & 0 & \sinh \xi
\\
\sinh \xi\sinh \psi & \cosh \psi & \cosh \xi\cosh\psi
\\
\cosh\psi\sinh\xi & \sinh\psi & \cosh\xi\cosh\psi
\end{pmatrix}.
\end{align}
\normalsize
In this case the decomposition formula \eqref{Decompo} can be used to obtain an elegant expression for the polynomials $R_{m,n}^{(\beta)}(i,k)$. For the parameter matrix \eqref{Matrix}, one has on the one hand
\begin{align}
\begin{aligned}
\label{CP-1}
&\BBraket{\beta}{i,k}{\mathcal{F}(\Lambda)}{m,n}{\beta}=\\
&\sqrt{\frac{(\beta)_{i+k}(\beta)_{m+n}}{i!k!m!n!}}
\left(\frac{\Lambda_{13}}{\Lambda_{33}}\right)^{i}
\left(\frac{\Lambda_{23}}{\Lambda_{33}}\right)^{k}
\left(\frac{-\Lambda_{31}}{\Lambda_{33}}\right)^{m}
\left(\frac{-\Lambda_{32}}{\Lambda_{33}}\right)^{n}
\left(\frac{1}{\Lambda_{33}}\right)^{\beta}
R_{m,n}^{(\beta)}(i,k),
\end{aligned}
\end{align}
and on the other hand
\begin{align}
\label{CP-2}
\BBraket{\beta}{i,k}{\mathcal{F}(\Lambda)}{m,n}{\beta}=\sum_{\mu,\nu\geqslant 0}\BBraket{\beta}{i,k}{\mathcal{F}(\Psi(\psi))}{\mu,\nu}{\beta}\BBraket{\beta}{\mu,\nu}{\mathcal{F}(\Xi(\xi))}{m,n}{\beta}.
\end{align}
Upon comparing the formulas \eqref{CP-1} and \eqref{CP-2} and using the one-parameter matrix elements \eqref{First} and \eqref{Second}, a direct computation shows that the parameters conspire to yield the expression
\begin{align}
\label{Tratnik-Type}
R_{m,n}^{(\beta)}(i,k)=\frac{(i+\beta)_{n}}{(\beta)_{n}}\,M_{m}(i;n+\beta;\tanh^2\xi)M_{n}(k;i+\beta;\tanh^2\psi).
\end{align}
The factorization \eqref{Tratnik-Type} of the bivariate Meixner polynomials as a product of two univariate Meixner polynomials is reminiscent of the bivariate Meixner polynomials defined by Tratnik. However, the Meixner polynomials \eqref{Tratnik-Type} do not exactly coincide with those defined in \cite{Tratnik-1991-04}. It is seen from \eqref{Rec-2} that in the special case \eqref{Matrix} one of the recurrence relations simplifies drastically. Indeed, \eqref{Rec-2} becomes
\begin{align*}
i\,R_{m,n}^{(\beta)}(i,k)&=\Big[m \cosh^2\xi+(m+n+\beta)\sinh^2\xi\Big]R_{m,n}^{(\beta)}(i,k)\nonumber
\\
&-\cosh^2\xi\,m\,R_{m-1,n}^{(\beta)}(i,k)-\sinh^2\xi\,(m+n+\beta)\,R_{m+1,n}^{(\beta)}(i,k)\nonumber,
\\
k\,R_{m,n}^{(\beta)}(i,k)&=\Big[m\sinh^2\xi\sinh^2\psi+n \cosh^2\psi+(m+n+\beta)\cosh^2\xi\sinh^2\psi\Big]R_{m,n}^{(\beta)}(i,k)\nonumber
\\
&+\sinh^2\psi\,m\,R_{m-1,n+1}^{(\beta)}(i,k)+\cosh^2\psi\sinh^2\xi\,n\,R_{m+1,n-1}^{(\beta)}(i,k)
\\
&-\cosh^2\xi\sinh^2\psi\;m\,R_{m-1,n}^{(\beta)}(i,k)-\sinh^2\xi\sinh^2\psi\,(m+n+\beta)\,R_{m+1,n}^{(\beta)}(i,k)\nonumber
\\
&-\cosh^2\xi\cosh^2\psi\;n\,R_{m,n-1}^{(\beta)}(i,k) -\sinh^2\psi\,(m+n+\beta)\,R_{m,n+1}^{(\beta)}(i,k).\nonumber
\end{align*}
The generating function \eqref{Generating} also has the simplification
\begin{align*}
\begin{aligned}
&(1-z_1-z_2)^{-\beta-i-k}(1-\coth^2\xi\,z_1)^{i}(1-z_1-\coth^2\psi\, z_2)^{k}
\\
&=\sum_{m,n\geqslant 0}\frac{(\beta)_{m+n}}{m!n!}\,R_{m,n}^{(\beta)}(i,k)\,z_1^{m}z_2^{n}.
\end{aligned}
\end{align*}
Note that polynomials \eqref{Tratnik-Type} corresponding to the special case \eqref{Matrix} are orthogonal with respect to the same weight function \eqref{Neg-Tri} as the generic polynomials.
\subsection{General case}
Let us now give a formula for the general bivariate Meixner polynomials. The most general $SO(2,1)$ pseudo-rotation can be taken of the form
\begin{align*}
\Lambda=R(\chi)\Psi(\psi)R(\theta),
\end{align*}
where $\Psi(\psi)$ is given by \eqref{Mat-a} and where $R(\theta)$, $R(\chi)$ are given by \eqref{Rotation}. For the matrix $\Lambda$, one has again on the one hand
\begin{align}
\begin{aligned}
\label{CP-3}
&\BBraket{\beta}{i,k}{\mathcal{F}(\Lambda)}{m,n}{\beta}=\\
&\sqrt{\frac{(\beta)_{i+k}(\beta)_{m+n}}{i!k!m!n!}}
\left(\frac{\Lambda_{13}}{\Lambda_{33}}\right)^{i}
\left(\frac{\Lambda_{23}}{\Lambda_{33}}\right)^{k}
\left(\frac{-\Lambda_{31}}{\Lambda_{33}}\right)^{m}
\left(\frac{-\Lambda_{32}}{\Lambda_{33}}\right)^{n}
\left(\frac{1}{\Lambda_{33}}\right)^{\beta}
R_{m,n}^{(\beta)}(i,k),
\end{aligned}
\end{align}
and on the other hand
\begin{align}
\begin{aligned}
\label{CP-4}
&\BBraket{\beta}{i,k}{\mathcal{F}(\Lambda)}{m,n}{\beta}=\\
&=\sum_{\mu,\nu,\rho,\sigma\geqslant 0}\BBraket{\beta}{i,k}{\mathcal{F}(R(\chi))}{\mu,\nu}{\beta}\BBraket{\beta}{\mu,\nu}{\mathcal{F}(\Psi(\psi))}{\rho,\sigma}{\beta}\BBraket{\beta}{\rho,\sigma}{\mathcal{F}(R(\theta))}{m,n}{\beta}.
\end{aligned}
\end{align}
Upon comparing the formulas \eqref{CP-3}, \eqref{CP-4} using the expressions \eqref{Second}, \eqref{Ultra} for the one-variable matrix elements, one arrives at the following formula for the general bivariate Meixner polynomials:
\small
\begin{align}
&R_{m,n}^{(\beta)}(i,k)=\nonumber\\
\label{Dompe-3}
&(-\tan^{2}\chi)^{k}(-\tan^2\theta)^{n}\sum_{\mu\geqslant0 }\frac{(-i-k)_{\mu}(-n-m)_{\mu}}{\mu!(\beta)_{\mu}}(\tan \chi \tan \theta \sinh \psi \tanh \psi)^{-\mu}\\
\nonumber
&\times K_{i+k-\mu}(k;\sin^2\chi;i+k)M_{m+n-\mu}(i+k-\mu;\mu+\beta;\tanh^2\psi)K_{n}(m+n-\mu;\sin^2\theta;m+n).
\end{align}
\normalsize
The formula \eqref{Dompe-3} thus gives an explicit expression of the bivariate Meixner polynomials in terms of the Krawtchouk and Meixner polynomials. It is directly seen that the summation appearing in \eqref{Dompe-3} is finite. Moreover, the duality property \eqref{Duality-2} of the bivariate Meixner polynomials is manifest in \eqref{Dompe-3} in view of the duality property of the univariate Krawtchouk and Meixner polynomials. Note that the comment below \eqref{Ultra} also applies for \eqref{Dompe-3}.
\section{Multivariate case}
In this section, it is shown how the results obtained thus far can easily be generalized to $d$ variables by considering the eigenspace of a bilinear expression in the creation/annihilation operators of $d+1$ harmonic oscillators.

Consider $d+1$ pairs of creation and annihilation operators $a_i^{\dagger}$, $a_{i}$ satisfying the Weyl algebra commutation relations
\begin{align*}
[a_i,a_k]=0,\qquad [a_i^{\dagger},a_{k}^{\dagger}]=0,\qquad [a_{i},a_{k}^{\dagger}]=\delta_{ik},
\end{align*}
for $i,k=1,\ldots,d+1$ and let $H$ be the Hermitian operator
\begin{align}
\label{Bilinear}
H=a_{1}^{\dagger}a_1+a_{2}^{\dagger}a_2+\cdots-a_{d+1}^{\dagger}a_{d+1}.
\end{align}
Let $\beta$ be a positive integer and denote by $\mathcal{V}_{\beta}$ be the infinite-dimensional eigenspace associated to the eigenvalue $1-\beta$ of $H$. An orthonormal basis for the space $\mathcal{V}_{\beta}$ is provided by the vectors
\begin{align}
\kket{n_1,\ldots,n_{d}}{\beta}=\ket{n_1,\ldots,n_{d},|n|+\beta-1},
\end{align}
where the notation $|n|=n_1+\ldots+n_{d}$ was used. The action of the operators $a_{i}^{\dagger}$, $a_i$ is identical to the one given in \eqref{Action-2}. Since \eqref{Bilinear} is clearly invariant under $SU(d,1)$ transformations, it follows that $\mathcal{V}_{\beta}$ provides a reducible representation space for the subgroup $SO(d,1)$. Again, one uses the notation $a_i=b_i$ for $i=1,\ldots,d$ and $a_{d+1}=b_{d+1}^{\dagger}$.

Let $B$ be a real $(d+1)\times (d+1)$ antisymmetric matrix and let $\Lambda$ be an orthochronous element of $SO(d,1)$. This means that $\Lambda$ satisfies
\begin{align*}
\Lambda^{t}\eta \Lambda=\eta,\qquad \Lambda_{d+1,d+1}\geqslant 1,
\end{align*}
where $\eta=\mathrm{diag}(1,1,\ldots,-1)$. Consider now the unitary representation
\begin{align}
\mathcal{F}(\Lambda)=\exp\left(\sum_{ij=1}^{d+1}B_{ij}b_{i}^{\dagger}b_{j}\right),
\end{align}
which has for parameters the $d(d+1)/2$ independent matrix elements of $B$. The transformations of the operators $b_i$, $b_i^{\dagger}$ under the action of $\mathcal{F}(\Lambda)$ are given by
\begin{align*}
\mathcal{F}(\Lambda)b_i\mathcal{F}^{\dagger}(\Lambda)=\sum_{k=1}^{d+1}\wt{\Lambda}_{ik}b_k,\qquad \mathcal{F}(\Lambda)b_i^{\dagger}\mathcal{F}^{\dagger}(\Lambda)=\sum_{k=1}^{d+1}\wt{\Lambda}_{ik}b_k^{\dagger},
\end{align*}
where $\wt{\Lambda}$ denotes the inverse matrix of $\Lambda$: $\wt{\Lambda}\Lambda=1$. Proceeding in as in Section 4, one can write the matrix elements of the reducible representations of $SO(d,1)$ on the space $\mathcal{V}_{\beta}$ as follows:
\begin{align*}
\BBraket{\beta}{x_1,\ldots,x_{d}}{\mathcal{F}(\Lambda)}{n_1,\ldots,n_{d}}{\beta}=W^{(\beta)}_{x_1,\ldots, x_{d}}\,M_{n_1,\ldots,n_{d}}^{(\beta)}(x_1,\ldots,x_{d}),
\end{align*}
with $M_{0,\ldots,0}^{(\beta)}(x_1,\ldots,x_{d})=1$ and where
\begin{align*}
W^{(\beta)}_{x_1,\ldots,x_d}=\BBraket{\beta}{x_1,\ldots,x_{d}}{\mathcal{F}(\Lambda)}{0,\ldots,0}{\beta}.
\end{align*}
By considering the identities $\BBraket{\beta+1}{x_1,\ldots,x_d}{\mathcal{F}(\Lambda)b_{i}}{0,\ldots,0}{\beta}=0$ for $i=1,\ldots,d$, one finds that $W_{x_1,\cdots,x_{d}}^{(\beta)}$ is given by
\begin{align*}
W_{x_1,\ldots,x_{d}}^{(\beta)}=\sqrt{\frac{(\beta)_{|x|}}{x_1!\cdots x_{d}!}}\,\Lambda_{1,d+1}^{x_1}\Lambda_{2,d+1}^{x_2}\cdots \Lambda_{d,d+1}^{x_d}\Lambda_{d+1,d+1}^{-\beta-|x|}.
\end{align*}
Since 
\begin{align}
\label{vlan}
\Lambda_{d+1,d+1}^2-\sum_{i=1}^{d}\Lambda_{d+1,i}^{2}=1,
\end{align}
one has
\begin{align}
\label{Neg-Multi}
|W_{i,k}^{(\beta)}|^2=w_{i,k}^{(\beta)}=\frac{(\beta)_{|x|}}{x_1!\cdots x_{d}!}\,(1-|c|)^{\beta}\,c_{1}^{x_1}\cdots c_{d}^{x_d},
\end{align}
with 
$$c_{i}=\frac{\Lambda_{i,d+1}^2}{\Lambda_{d+1,d+1}^{2}},\qquad i=1,\ldots,d
$$
In view of \eqref{vlan}, the condition $|c|<1$ is identically satisfied and hence the following normalization condition holds:
\begin{align*}
\sum_{x_1,\ldots,x_{d}\geqslant 0}w_{x_1,\ldots,x_{d}}^{(\beta)}=1.
\end{align*}
The raising relations \eqref{Raising} are readily generalized to $d$ variables and from there it is seen that $M_{n_1,\ldots,n_{d}}^{(\beta)}(x_1,\ldots,x_{d})$ are polynomials in the variables $x_1,\ldots,x_{d}$ of total degree $|n|$. As a consequence of the unitarity of the operator $\mathcal{F}(\Lambda)$, the polynomials $M_{n_1,\ldots,n_{d}}^{(\beta)}(x_1,\ldots,x_{d})$ satisfy the orthogonality relation
\begin{align}
\sum_{x_1,\ldots,x_{d}\geqslant 0}w_{x_1,\ldots,x_{d}}^{(\beta)}\,M_{n_1,\ldots,n_{d}}^{(\beta)}(x_1,\ldots,x_{d})\,M_{m_1,\ldots,m_{d}}^{(\beta)}(x_1,\ldots,x_{d})=\delta_{n_1,m_1}\cdots \delta_{n_d,m_{d}},
\end{align}
with respect to the negative multinomial distribution \eqref{Neg-Multi}. The calculation of the generating function of section 5 is also easily generalized to an arbitrary finite number of variables. One then obtains the generating function used by Griffiths and Iliev to define the $d$-variable Meixner polynomials $R_{n_1,\ldots,n_{d}}^{(\beta)}(x_1,\ldots,x_{d})$:
\begin{align*}
&(1-|z|)^{-\beta-|x|}\prod_{i=1}^{d}\left(1-\sum_{j=1}^{d}u_{i,j}z_j\right)^{x_i}
\\
&=\sum_{n_1,\ldots,n_{d}\geqslant 0}\frac{(\beta)_{|n|}}{n_1!\cdots n_{d}!}\,R_{n_1,\ldots,n_{d}}^{(\beta)}(x_1,\ldots,x_{d})\,z_1^{n_1}\cdots z_{d}^{n_{d}}.
\end{align*}
where the parameters $u_{i,j}$ are given by
\begin{align*}
u_{i,j}=\frac{\Lambda_{i,j}\Lambda_{d+1,d+1}}{\Lambda_{i,d+1}\Lambda_{d+1,j}},
\end{align*}
for $i,j=1,\ldots,d$. All properties of the multivariate Meixner polynomials can be derived in complete analogy with the $d=2$ case which has been treated in detail here.
\section{Conclusion}
In summary, we have considered the reducible representations of the $SO(d,1)$ group on the eigenspace of a bilinear expression in the creation/annihilation operators of $d+1$ independent quantum harmonic oscillators. We have shown that the multivariate Meixner polynomials arise as matrix elements of these $SO(d,1)$ representations and we have seen that the main properties of the polynomials can be derived systematically using the group theoretical interpretation. 

In \cite{Miki-2012-03}, the bivariate Krawtchouk polynomials were seen to occur as wavefunctions of a Hamiltonian describing a discrete/finite model of the harmonic oscillator in two dimensions possessing a $SU(2)$ symmetry. This result and the considerations of the present paper suggest that the bivariate Meixner polynomials could also arise as wavefunctions of a discrete Hamiltonian. We hope to report on this issue in the future.
\section*{Acknowledgments}
The authors wish to thank Robert Griffiths for stimulating discussions while he was visiting the Centre de Recherches Math\'ematiques as Clay Senior Scholar. V.X.G holds an Alexander-Graham-Bell fellowship for the Natural Sciences and Engineering Research Council of Canada (NSERC). The research of L.V. is supported in part by NSERC.
\section*{References}

\end{document}